\documentclass{aa}
\usepackage{natbib}
\usepackage{graphicx}
\usepackage{times}
\usepackage{amsmath}
\usepackage{txfonts}
\newcommand{\MWC}{MWC\,480}
\newcommand{\LkCa}{LkCa\,15}
\newcommand{\msun}{{\,\rm M}_{\odot}}

\newcommand{\nht}{\ifmmode {{\rm NH}_3} \else {NH{\bas 3}} \fi}
\newcommand{\tco}{\ifmmode {^{13}{\rm CO}} \else {$^{13}{\rm CO}$}\fi}
\newcommand{\dco}{\ifmmode {^{12}{\rm CO}} \else {$^{12}{\rm CO}$}\fi}
\newcommand{\cdo}{\ifmmode {{\rm C}^{18}{\rm O}} \else {${\rm C}^{18}{\rm
O}$}\fi}

\newcommand{\htco}{\ifmmode {{\rm H}^{13}{\rm CO}^{+} } \else {${\rm H}^{13}
{\rm CO}^{+}$ }\fi}
\newcommand{\hco}{\ifmmode {{\rm H}^{12}{\rm CO}^{+} } \else {${\rm H}^{12}
{\rm CO}^{+}$ }\fi}
\newcommand{\ndhp}{\ifmmode {{\rm N}_{2}{\rm H}^{+} } \else {${\rm N}_{2}
{\rm H}^{+}$ }\fi}
\newcommand{\juz}{\ifmmode {{\rm J}=1\rightarrow 0} \else
{J=1$\rightarrow$0}\fi}
\newcommand{\jdu}{\ifmmode {{\rm J}=2\rightarrow 1} \else
{J=2$\rightarrow$1}\fi}
\newcommand{\jtd}{\ifmmode {{\rm J}=3\!\rightarrow\!2} \else
{${\rm J}=3\!\rightarrow\!2$} \fi}
\newcommand{\jcq}{\ifmmode {{\rm J}=5\!\rightarrow\!4} \else
{${\rm J}=5\!\rightarrow\!4$} \fi}
\newcommand{\as}{\ifmmode {^{\scriptscriptstyle\prime\prime}}
        \else $^{\scriptscriptstyle\prime\prime}$\fi}
\newcommand{\am}{\ifmmode {^{\scriptscriptstyle\prime}}
        \else $^{\scriptscriptstyle\prime}$\fi}
\newcommand{\TableGEOM}{
\begin{table}
\caption{Comparison of the inclinations and orientations of the \LkCa\ and \MWC\ disks derived from CO
isotopologues \citep[see][]{Pietu_etal2006} and mm continuum emission. \label{tab:geom} }
\begin{tabular}{lcc}
 \hline \hline
 & from CO & from 2.8 and 1.4 mm  \\
 & isotopologues &  emission \\
 \hline
        \multicolumn{3}{c}{\LkCa} \\
 PA ($^\circ$) & $150 \pm 1$ & $151 \pm 3$    \\
 $i$ ($^\circ$) & $52 \pm 1$ & $49 \pm 3$ \\
 \hline
         \multicolumn{3}{c}{\MWC} \\
 PA ($^\circ$) & $58 \pm 1$   & $61 \pm 8$ \\
 $i$ ($^\circ$) & $36 \pm 1$ & $31 \pm 5$
 \end{tabular}\\
\end{table}
}

\newcommand{\NEWTableLkCa}{
\begin{table}
\caption{ Parameters of the \LkCa\ disk derived from the model fits. \label{tab:fit-lkca15} }
\begin{tabular}{lcc}
\hline \hline
\multicolumn{3}{c}{Temperature from CO isotopologues} \\
 $T_{100}$ (K) & \multicolumn{2}{c}{$22 \pm 1$} \\
 $q$ &  \multicolumn{2}{c}{$0.37 \pm 0.02 $} \\
 \hline
  \multicolumn{3}{c}{Derived parameters from 2.8 and 1.4 mm dust emission} \\
 $R_\mathrm{int}$ (AU) & $46 \pm 3$ & [5]$^a$ \\
 $R_\mathrm{out}$ (AU) & $177 \pm 12$ & $135 \pm 3$ \\
 $\Sigma_{100}$ (g.cm$^{-2}$) & $3.1\pm 0.4$ & $4.1 \pm 0.6$ \\
 $p$ & $1.7 \pm 0.3$ & $-0.5 \pm 0.2$  \\
 $\beta$ & $1.2 \pm 0.1$ & $1.1 \pm 0.1$  \\
 $M_\mathrm{disk}$ ($\msun$) & 0.029 & 0.025 \\
\end{tabular}\\
{The error bars are the $1\sigma$ formal errors (effective noise, see Sect.2) from the fit. Two solutions are
presented: one with a free inner radius, and one with a fixed inner radius of 5 AU. The dust emission is too
optically thin to allow us an independent fit of the temperature, which is adopted from the CO analysis
presented in \citet{Pietu_etal2006}. (a) Square brackets indicate fixed parameters.}
\end{table}
 }
\newcommand{\NEWTableMWC}{%
\begin{table}
\caption{Parameters of the \MWC\ disk derived from the model fits.\label{tab:fit-mwc480}}
\begin{tabular}{lcc}
\hline \hline
\multicolumn{3}{c}{Fixed input parameters} \\
 $R_\mathrm{int}$ (AU) & [3]$^a$  &  [3] \\
 $q$ & [0] & [0.5] \\
 \hline
   \multicolumn{3}{c}{Derived parameters from 2.8 and 1.4 mm dust emission} \\
 $T_{100}$ (K) & $24.5 \pm 3$ & $9.8 \pm 0.4$ \\
 $R_\mathrm{out}$ (AU) &  $190 \pm 15 $ & $ 185 \pm 15 $\\
 $\Sigma_{100}$ (g.cm$^{-2}$) & $3.1 \pm 0.2$ & $ 17.1 \pm 1.5$ \\
 $p$  & $2.5 \pm 0.1$ & $1.6 \pm 0.1$ \\
 $\beta$ & $1.26 \pm 0.05$ & $1.51 \pm 0.06$ \\
 $M_\mathrm{disk}$ ($\msun$) & 0.24 & 0.30 \\
 $M (R>35\,\mathrm{AU})$ ($\msun$) & 0.04 & 0.19 \\
\end{tabular}\\
{The error bars are the $1\sigma$ formal errors (effective noise, see Sect.2) from the fit. $R_{int}$ has
been fixed to 3 AU. (a) Square brackets indicate fixed parameters.}
\end{table}
}

\newcommand{\FigAll}{%
  \begin{figure*}
    \centering %
    \includegraphics[angle=270,width=18.0cm]{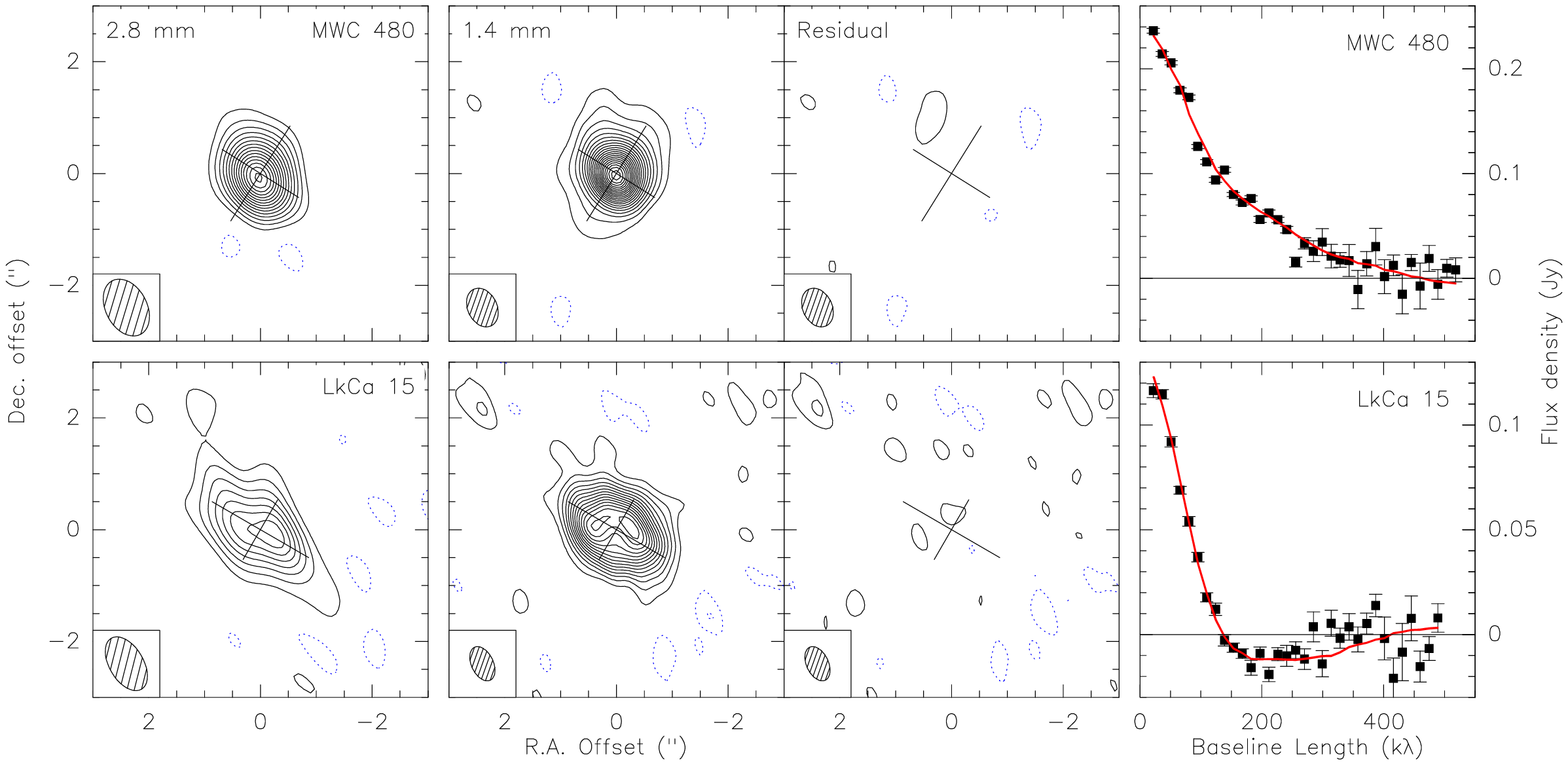}
    \caption{Top row: Results for \MWC. From left to right: 2.8 mm
    continuum image.  The angular resolution is $1.09 \times 0.74$ at
    PA $28^\circ$, and the contour spacing is 1 mJy/beam (0.13 K,
    about 3.5 $\sigma$). The 1.4 mm continuum image: the angular
    resolution is $0.73 \times 0.53$ at PA $33^\circ$, and the contour
    spacing is 5 mJy/beam (0.33 K, $2.5 \sigma$). The 1.4 mm residual
    from the best fit with the same contours.  Right: 1.4\,mm real
    part of the visibility $vs$ baseline lengths. The points and error
    bars represent the measured values, and the curve indicates the
    visibilities of the best-fit disk model. The visibilities have
    been corrected from the disk orientation and inclination by
    compressing the V value by $\cos(i)$ prior to circular averaging.
    Bottom row: Results for \LkCa. The 2.8 mm image is obtained by
    considering baselines longer than 200 m (but no flux was lost in
    this process). The resolution is $1.06 \times 0.61$ at PA
    $31^\circ$, and the contour spacing is 0.75 mJy/beam (0.12 K, 2
    $\sigma$). The 1.4 mm resolution is $0.66 \times 0.39$ at PA
    $55^\circ$, and the contour spacing is 1.5 mJy/beam (or 0.15 K,
    1.7 $\sigma$).  The crosses indicate the position, orientation, and
    aspect ratio of the disks.}
  \label{fig:all}
  \end{figure*}}

\begin{document}
%---------------------------------------
\title{Resolving the inner dust disks surrounding \LkCa\ and \MWC\ at mm wavelengths
\thanks{Based on observations carried out with the IRAM Plateau de
Bure Interferometer. IRAM is supported by INSU/CNRS (France), MPG (Germany), and IGN (Spain).}}

\author{Vincent Pi\'etu \inst{1}, Anne Dutrey \inst{2}, St\'ephane
Guilloteau \inst{2}, Edwige Chapillon \inst{1}, and Jer\^ome Pety
\inst{1}}
\offprints{Vincent Pi\'etu, \email{pietu@iram.fr}}
\institute{IRAM, 300 rue de la piscine, F-38406 Saint Martin d'H\`eres, France \and L3AB, CNRS UMR5804, OASU,
2 rue de l'Observatoire, BP 89, F-33270 Floirac, France}
\date{Received / Accepted}
%-----------------------------
\abstract{}
{We constrain the dust distribution and its properties (temperature, emissivity) in inner proto-planetary
disks}
{We performed sub-arcsecond high-sensitivity interferometric
  observations of the thermal dust emission at 1.4\,mm and 2.8\,mm in
  the disks surrounding \LkCa\ and \MWC , with the new 750 m baselines
  of the IRAM PdBI array. This provides a linear resolution of $\sim
  60$ AU at the distance of Taurus.}
{We report the existence of a cavity of $\sim 50$ AU radius in the
inner disk of \LkCa. Whereas \LkCa\ emission is optically thin, the
optically thick core of \MWC\ is resolved at 1.4 mm with a radius of
$\sim 35$ AU, constraining the dust temperature. In \MWC, the dust
emission is coming from a colder layer than the CO emission, most
likely the disk mid-plane.}
{These observations provide direct evidence of an inner cavity around \LkCa. Such a cavity most probably
results from the tidal disturbance created by a low-mass companion or large planet at $\sim 30$ AU from the
star. These results suggest that planetary system formation is already at work in \LkCa. They also indicate
that the classical steady-state viscous disk model is too simplistic a description of the inner 50 AU of
``proto-planetary'' disks and that the disk evolution is coupled to the planet formation process. The \MWC\
results indicate that a proper estimate of the dust temperature and size of the optically thick core are
essential for determining the dust emissivity index $\beta$.}
%
%-------------------------------------------------
\authorrunning{Pi\'etu et al., }
\titlerunning{Resolving the \LkCa\ and \MWC\ inner disks}
\maketitle
%-------------------------------------------------

\FigAll

\section{Introduction}

With current instrumentation, proto-planets located at the distance of Taurus cannot be observed directly.
The best evidence of planet formation in proto-planetary disks remains the detection of the tidal gaps.
Because of the dependence of the dust temperature on the distance from the star, these gaps should leave
signatures in the infrared SED of the objects, in the form of a deficit of emission at wavelengths between
$\sim$ 1 and 30 $\mu$m. \citet{Koerner_etal1993} suggest that the deficit of emission at 10 $\mu$m in GM Aur
was indeed the result of such a gap. Many recent studies of the spectral energy distribution (SED) in the
near-infrared (NIR) and mid-infrared (MIR) coming from data obtained by various telescopes, in particular
from the Spitzer satellite, have revealed NIR-MIR emission dips that can be interpreted by truncated disks
with inner radii of $\sim 3-5$ AU up to $\sim$ 25 AU \citep{Calvet_etal2005}. Some direct evidence of inner
cavities has been also obtained in the scattered light images of the disks of moderate opacity associated
with Herbig Ae stars such as HD141569 \citep{Augereau_etal1999}. More recently, a large inner radius (70 --
100 AU) was measured in the mm domain for the dense spiral-like structure found in AB Auriga
\citep{Pietu_etal2005}.

However, models of planet formation predict gaps with density contrasts of only $\sim$ 10-100 for a
Jupiter-like planet \citep[e.g][]{Crida_etal2006}. Due to the very high dust opacity observed in the IR (at
$\sim 1 \mu$m, assuming standard dust grain properties, the opacity can be as high as $10^4 -10^6$, see
\citealt{dAlessio_etal1998}), the IR SED does not allow astronomers to get detailed knowledge on the moderate
density contrasts. This can be qualitatively understood in Fig.5 in \citet{Dullemond_etal2001}, where a
change in the slope $p$~of 1 on the dust surface density does not significantly affect the SED (by less than
$\sim$ 20~\%), which is therefore difficult to disentangle from the other physical effects on real data.

As a consequence, only gaps corresponding to a significant decrease in the surface density in the inner disk,
i.e. a transition in opacity between thick and thin regimes, can be detected in the modelling of the IR SEDs.
Only resolved images obtained with a tracer of moderate opacity can provide reliable insights into the
geometrical structure and density contrasts when the inner cavities are not significantly empty. The dust
opacity of proto-planetary disks in the millimeter/sub-millimeter range is well adapted to imaging such
``young'' cavities, provided enough angular resolution and sensitivity is available.

In this paper, we present new PdBI observations at an angular resolution of $0.35''-0.5''$ for the Keplerian
disks surrounding two bona-fide single PMS stars \LkCa\ and \MWC. \LkCa\ is a classical TTauri star of
spectral type K5, mass $\sim 1\msun$, and age $\sim 3-5$~Myr and \MWC\ is a Herbig Ae star of spectral type
A4 and age $\sim 5-7$~Myr \citep{Simon_etal2000}.

\section{Observations and results}

The observations were performed from winter 2001 to winter 2005.  We observed simultaneously at 110 and 220
GHz (2.8 and 1.4\,mm).  Typical system temperatures range from 200 to 400 K at 220 GHz. We mainly used 4
configurations: the standard BCD configuration of the IRAM array, with (projected) baselines ranging from 15
m to 320 m, and the new A$^+$ configuration, which provides baselines extending up to $\sim$750~m. Some
additional data were obtained on MWC 480 in A configuration, using time sharing with AB Aur
\citep[see][]{Pietu_etal2005}. The data were obtained under good weather conditions: the rms phase did not
exceed $\sim 70^\circ$ at 1.4\,mm even on the longest baselines. The two sources were observed in
track-sharing mode, alternating every 10 minutes from one source to the other. For the BCD configurations, we
calibrated the data using the standard method, with MWC 349 as a reference for flux calibration. On the
longest baselines, MWC 349 is partially resolved at 1.4 mm, and a different method was required for the flux
calibration. We first based our flux scale on an extrapolation in time of the flux densities of the two phase
calibrators, 0415+379 (i.e. 3C111) and 0528+134. However, since quasars have unpredictable flares, we also
compared the measured flux density of \MWC\ obtained from the BCD data to that obtained in the A$^+$ data on
the overlapping baseline range (from 100 to 400 m). \MWC\ is bright and compact enough to allow such a direct
comparison, to a precision of $\sim 10$ \%. The two methods give the same results within 5 \%. The
simultaneous observation and common calibration ensure that any morphological or intensity difference
observed between the two sources is real and not due to instrumental artifacts.

At 220 GHz, these observations provide an angular resolution about $0.7 \times 0.4''$ using natural
weighting. With uniform weighting, the longest baselines (500 k$\lambda$) allow us to reach an angular
resolution of about $0.5 \times 0.3''$ at PA $\sim 45^\circ$. At  110 GHz, the resolution is a factor lower.
Based on the integration time, system noise, and measured efficiencies, the expected (thermal) noise level
was 0.7 mJy/beam at 220 GHz. However, the dynamic range is limited by phase noise. This results in an
effective noise of 0.9 mJy/beam for \LkCa{} and 2.0 mJy/beam for \MWC. At 110 GHz, the noise is 0.3 mJy/beam,
so essentially thermal.

Figure \ref{fig:all} presents the images obtained at 2.8\,mm and 1.4\,mm on \MWC\ and \LkCa. While the disk
of \MWC\ appears centrally peaked, the disk around \LkCa\ does not. The 1.4\,mm image of \LkCa\ is rather
similar to the first image of GG Tau \citep[ see their Fig.1]{Dutrey_etal1994} where the inner radius of the
circumbinary dust ring was just resolved by the interferometer. The total flux densities at 2.8 and 1.4 mm
are $17.0 \pm 0.8$~mJy and $140\pm 3$~mJy, respectively, for \LkCa, corresponding to an apparent spectral
index of $\alpha = 3.04 \pm 0.07$. For \MWC, S$(2.8) = 35.2 \pm 0.8$~mJy and S$(1.4) = 235 \pm 4$~mJy,
leading to $\alpha = 2.74 \pm 0.04$. The calibration uncertainties result in an additional error of $\pm0.15$
on the absolute values of $\alpha$, but the relative values are not affected, since the data were obtained
simultaneously. Figure \ref{fig:all} also shows the residual images after subtraction of the best-fit model
(see Tables \ref{tab:fit-mwc480}-\ref{tab:fit-lkca15}). The rightmost panels of Fig.\ref{fig:all} display the
real part of circular average of the calibrated visibilities obtained after deprojection, superimposed with
the visibilities from the best models (curves). \LkCa\ is well resolved with a first null at 130 k$\lambda$
(and probably a second null at $\sim$ 400~k$\lambda$).

\section{Data analysis}

\TableGEOM

\NEWTableMWC

To better quantify the obvious morphologic differences between both sources, the 2.8\,mm and 1.4\,mm data
were analyzed in the Fourier Plane using our disk fitting method \citep[see][ and references
therein]{Dutrey_etal2006}.  We modelled an inclined disk with truncated inner and outer radius and power law
distribution of the temperature $T(r) = T_{100} (r/100\,\mathrm{AU})^{-q}$, surface density $\Sigma(r) =
\Sigma_{100} (r/100\,\mathrm{AU})^{-p}$. Following \citet{Beckwith_etal1990}, we used the
\citet{Hildebrand_1983} prescription for the dust opacity, $\kappa(\nu) = 0.1(\nu/1000\,\mathrm{GHz})^\beta$
cm$^2$/g (per gram of dust+gas). We simultaneously fit the 2.8 and 1.4 mm UV data, thereby constraining
$\beta$ directly. As shown in Table \ref{tab:geom} for both sources, the inclinations and orientation derived
from the continuum data agree with those determined from an analysis of the $^{12}$CO \citep{Simon_etal2000}
and $^{13}$CO observations \citep[for details, see][]{Pietu_etal2006}. We used the more accurate CO-based
inclinations in the analysis. For \MWC\ the temperature can be constrained by the dust observations. For
\LkCa, we used the temperature determined by \citet{Pietu_etal2006} from CO lines. We checked that the choice
of the scale height does not influence the results at all: we used $h(r) = 16.5 (r/100\,\mathrm{AU})^{1.25}$
AU for both sources.

\NEWTableLkCa

For continuum data, the brightness distribution scales in the optically thick case as the temperature $T$,
and its exponent $q$, while in the optically thin case, it scales as the product $\Sigma T$ and the sum
$p+q$. Since in general the dust opacity decreases with radius, one can measure both the surface density and
the temperature with angular resolution that is high enough.

Table \ref{tab:fit-mwc480} shows the best fits for \MWC. The large value of the peak surface brightness (7 K
in Fig.\ref{fig:all}) indicates a significant opacity at both wavelengths in the inner $\sim 35$ AU.  Indeed,
optical depth reaches unity between $r=31$ AU (model 1) and 38 AU (model 2) at 220 GHz, and between 20 and 23
AU at 110 GHz. We thus attempted to treat the dust temperature as a free parameter. The measurements directly
constrain the temperature around $20 - 30$~AU to be $\sim 20$~K. However, the optically thick core is not
sufficiently resolved to determine the temperature exponent $q$. Table \ref{tab:fit-mwc480} shows two
solutions. The data are compatible with a $T_{100}=24$~K, $q=0$ or with steeper laws, $T_{100}=10$~K,
$q=0.5$, but not with higher temperatures.
The temperature derived from $^{12}$CO ($T_{100}=45$~K and $q=0.65$) is much too high to apply to the dust.
All solutions intersect near $r=17$~AU with $T = 24$~K. The lack of knowledge on the temperature law results
in a significant uncertainty in the surface density law, and thus on the total disk mass. With these
opacities, the optically thick inner core contributes quite significantly to the total flux, and the dust
emissivity index $\beta = 1.4\pm0.15$ is thus substantially higher than $\alpha-2$.

For \LkCa, the surface brightness of the mm emission is low ($<3$~K), indicating that the dust is essentially
optically thin. It is thus impossible to constrain the dust temperature with this data. On the other hand, an
analysis of \dco\ and \tco\ by \citet{Pietu_etal2006} provides a unique value for the temperature, $T_{100} =
22 \pm 1$ and $q = 0.37 \pm 0.02$. We adopted these values.  With a small inner radius (5 AU, but the result
do not change for lower values), the best-fit solution implies $p \simeq - 0.5$, i.e. a surface density
increasing with radius. This is physically implausible. Treating the inner radius as a free parameter results
in a better fit at the 4 $\sigma$ level and a more physical value for $p$ (see Table \ref{tab:fit-lkca15}).
Other combinations of inner, outer radius and density exponent are possible: for example, a solution with $p
\simeq 3$, $R_\mathrm{int} \simeq 65$ AU and $R_\mathrm{out} > 600$ AU (in agreement with the disk extent in
CO) is only 2 -- 3 $\sigma$ above the best-fit solution for \LkCa. This may indicate that \LkCa\ is not
devoid of dust beyond $\sim 170$ AU, but surrounded by a more tenuous disk. Despite the correlation between
$p$ and $R_\mathrm{in}$, the \LkCa\ disk exhibits a clear decrease in surface density towards its center,
with the best solution indicating a $\simeq 50$ AU radius central hole.

In conclusion, while \MWC\ presents a ``classical'' dust distribution, \LkCa\ indicates that dust disks can
present brightness distributions versus radius that are more sophisticated than simple power laws, and
suggests the existence of a large inner cavity in an otherwise ``standard'' dust disk.

\section{Discussion}

\subsection{\MWC}

\MWC\ was recently observed at 1.4\,mm by \citet[][ hereafter HLM]{Hamidouche_etal2006} using the BIMA array,
with slightly higher resolution, but lower sensitivity. HLM mention a similar source orientation. They use a
temperature derived from the IR SED, $T_{100} = 17$~K, and $q=0.62$. However, because of the IR opacity, the
IR-constrained temperature deduced from an isothermal vertical disk model naturally overestimates the
mid-plane temperature to which the mm emission is sensitive \citep[see also][ for
details]{dAlessio_etal1998}. Indeed, the temperature used by HLM is much too high to explain our
observations, and using it results in a best-fit solution that is $11 \sigma$ worse than those presented in
Table \ref{tab:fit-mwc480}. HLM also derive a relatively shallow surface density distribution, $p = 0.5 -
1.0$, as a result of the high assumed temperature. The HLM temperature law requires a low opacity to
reproduce the surface brightness (9~K in their Fig.1).  As a consequence, the shallow
inner-surface-brightness distribution is represented by a flat surface density, while it is in reality due to
contribution by the optically thick core. Also, $p$ is biased towards low values by the extended emission
present in the HLM image, which is not confirmed in ours. This structure may be due to phase errors. Note
that in general, phase noise will result in a flattening of the brightness distribution, since it scatters
signal (much like an effective seeing).

The value of $\beta = 1.4 \pm 0.15$ indicates moderate grain growth. This is similar to the value obtained
for AB Aur by \citet{Pietu_etal2005}. Using the coronagraphic mode of the HST/NICMOS camera at $1.6\,\mu$m,
\citet{Augereau_etal2001} observed the disk of \MWC, but did not detect its scattered light. Their upper
limit on the column density of scattering material is $\sim 10^{-4}$g.cm$^{-2}$ at 100~AU (of dust only),
several orders of magnitude below our measured value.  Our results suggest that a significant amount of dust
is still ``hidden'' to the scattered light regime, most likely because the dust disk is geometrically thin
and the starlight partially masked by the dust orbiting close to the star. This can occur if the dust has
significantly settled on the disk-mid plane. This explanation agrees with the classification of \MWC\ as a
group II star \citep{Meeus_etal2001,Acke_etal2004}. The constraint on the dust temperature indicates that the
dust is colder than the zone traced by the CO isotopologues, in further agreement with this hypothesis.
Although indirect, these arguments suggest that the dust has settled towards the disk plane.  Table
\ref{tab:fit-mwc480} indicates the \MWC\ disk is quite massive. However, this is largely the result of
extrapolation of the density and/or temperature laws, as indicated by the total mass beyond 35 AU in Table
\ref{tab:fit-mwc480}.

\MWC\ illustrates the need to resolve the inner optically thick core and to \textbf{measure} the dust
temperature to properly characterize the dust emissivity. Lower-resolution data \citep{Dutrey_etal1996}
typically tend to find values of $\beta \simeq 1$, and our new results indicate such values could be
significantly biased by contamination from optically thick dust at low temperature.

\subsection{The inner disk of \LkCa}

Until now, observations from the NIR to the mm range have shown that \LkCa\ is a bona-fide Classical TTauri
star surrounded by a large CO disk of a few $0.01 \msun$ \citep{Simon_etal2000,Qi_etal2003}.
\citet{Leinert_etal1993} observed \LkCa\ and classified it as a single star, since they found no companion
down to a separation of $\sim 0.13''$ (20 AU).

\citet{Bergin_etal2004} have observed the H$_2$ UV emission and the NIR-MIR SED of \LkCa. They conclude that
the disk has an inner radius (puffed-up inner rim) located at $\sim 3$~AU. From their disk modelling, they
estimate a column density of dust $\Sigma_\mathrm{dust} \simeq 0.2$g.cm$^{-2}$ at 1 AU, assuming a standard
accretion disk model with $\alpha = 0.01$ and $T=100$~K. Extrapolated to 45 AU, this translates into 0.07
g.cm$^{-2}$ of dust+gas (with a gas-to-dust ratio of 100). This is 200 times lower than the surface density
from Table~\ref{tab:fit-mwc480}, confirming the interpretation of the mm images of \LkCa\ as the result of a
cavity of radius $\sim 45$ AU nearly devoid of dust. If we fit a $p=1$ surface density law into the \LkCa\
cavity, we find an upper limit of $\Sigma_\mathrm{dust} \simeq 0.12$ g.cm$^{-2}$ at 1 AU, consistent with the
results of \citet{Bergin_etal2004}.

In the current knowledge of inner-disk destruction processes, among the mechanisms (such as planetary
formation, photo-evaporation, radiation pressure, etc\ldots) that can explain this distribution, planetary
formation or a low mass companion remain plausible since the \LkCa\ disk is fairly massive. The large disk
density makes photo-evaporation and radiation pressure unlikely to be effective. \citet{Alexander_etal2006}
showed that when photo-evaporation starts propagating beyond 20 -- 30 AU, the surface density near 50 AU is
$\simeq 0.03$ g.cm$^{-2}$, within a factor of a few to account for different UV fluxes (see their Fig.1).
This is 500 times lower than derived from our best fit model in Table \ref{tab:fit-lkca15}.

In a viscous disk, a gap can be opened by a planet as soon as the Hill's radius is larger than the disk
height. Following \citet{Takeuchi_etal1996}, the half-width of a gap created by a proto-planet or a low-mass
companion can be approximated by: $w = 1.3 a A^{1/3}$ where $a$ is the semi-major axis of the orbit. The
strength ratio of tidal to viscous effects  $A$ is :
$$A = (M_p/ M_*)^2 \frac{1}{3\alpha(h(r)/r)^2}$$ For $a \simeq 30$~AU
and assuming $\alpha \sim 0.01$, $h(r)/r \simeq 0.1$, $M_* \simeq 1 \msun$ and the mass of the planet $M_p
\simeq 0.005 \msun$, we find $A \simeq 0.083$ and $w \simeq 17$ AU. Hence a $\sim 5-10$ Jupiter mass planet
orbiting at 30 AU would be sufficient to evacuate the inner 50 AU of the \LkCa\ disk.

Getting an upper limit on the orbiting object mass is more difficult. The separation limit from
\citet{Leinert_etal1993} is 20 AU, but given the inclination of the \LkCa\ disk, an object orbiting at 30 AU
would spend most of its time at projected distances about 20 AU.  The best limit on the companion mass comes
from the kinematic determination of \citet{Simon_etal2000}, who indicate a total mass for the system of
$1.0\pm0.1 \msun$. The spectral type of \LkCa\ makes it unlikely to be a star of less than $0.8 \msun$,
leaving at most $0.2 \msun$ for a companion. A coeval star of this mass would have a K magnitude of about
10.5.

\section{Summary}

We used the new 750\,m baselines of the IRAM array to image the dust disks orbiting \MWC\ and \LkCa\ at an
angular resolution of $\sim 0.4''$.
\begin{itemize}\itemsep 0pt
 \item The dust disk of \MWC\ is centrally peaked with an optically
 thick core of radius $\sim 35$~AU at 1.4\,mm. Since the core is
 resolved, we can estimate its temperature, which is significantly
 lower than that of the CO layers, and measure the dust emissivity
 index $\beta$.  The low dust temperature suggests that grains
 emitting at mm wavelengths tend to be located around the
 colder disk mid-plane, in agreement with sedimentation.  The index
 $\beta$ is $\sim 1.4$, suggesting moderate grain growth, like in AB
 Aur.

\item To be compatible with existing NIR images of \MWC{}, the dust must be confined to a geometrically thin
disk and mostly hidden from the starlight.

\item The \LkCa\ disk reveals a large ($\sim $50 AU radius) cavity, which is not completely devoid of dust.
The most probable explanation for such a wide cavity is the existence of a massive planet ($> 0.005 \msun$)
or of a low mass companion ($<0.2 \msun$) orbiting around $\sim 30$~AU from the star.
\end{itemize}
Long baselines on large mm/submm arrays are now opening the planetary formation regions of nearby young disks
to investigations. However, as for AB Aur, figuring out whether the companion to \LkCa\ is a proto-planet or
a low-mass star requires deeper IR images, or the advent of ALMA, which may resolve the structure of the
cavity.

\begin{acknowledgements} We thank the referee for very useful comments.
We thank A.Morbidelli and A.Crida for a very helpful discussion of the formation of gaps in disk. We
acknowledge M.Simon for discussion of the binarity of \LkCa\ and J.-C.Augereau for considerations of the
scattered light image of \MWC. We thank the Plateau de Bure IRAM staff, especially J.-M.Winters, for their
help during the observations. This research was supported by the French Program ``Physique Chimie du Milieu
Interstellaire'' (PCMI).
\end{acknowledgements}

%
% Warning!
%\bibliography{mybib}
\bibliography{inner1}
\bibliographystyle{aa}
\end{document}